\documentstyle[aps,multicol,epsf]{revtex}

\begin{document}

\begin{multicols}{2}

\narrowtext

\noindent 
{\bf Comment on ``Breakdown of the Internet under Intentional Attack''} \\

In a recent paper \cite{cebh01}, Cohen, Erez, ben-Avraham, and Havlin studied the problem of intentional attack in equilibrium scale-free networks  
\cite{ajb00}. Their study focused on the ``exact value of the critical fraction'' $p_c$ of the number of sites needed for disruption \cite{cebh01}, and they have found that such networks are highly sensitive to sabotage of a small fraction of the total number of sites. 
However, in Ref.  \cite{cebh01}, a continuum approximation for the scale-free degree distribution $P(k)$ of the undamaged network was actually used. Its results were found to agree with the results of the 
simulation  
\cite{cebh01} 
for the transformed degree distribution $\int_{k+1/2}^{k-1/2}dqP(q)$ ``rendering the continuum approximation'' but no natural comparison with  
results following from  
the original discrete distribution $P(k)$ was made. 
Here we obtain the exact values of $p_c$ 
using the genuine discrete degree distribution $P(k)$. These values agree with the results of Ref. \cite{cnsw00} and demonstrate that Ref. \cite{cebh01} seriously overestimates the resilience of scale-free networks to intentional damage.      

The exact results can be easily obtained using the ideas of Ref. \cite{cebh01}.  
The intentional damage cuts off the sites with $k > \tilde{K}(p)$ where the cutoff $\tilde{K}(p)$ can be obtained from the relation
$p=1-\sum_{k=0}^{\tilde{K}(p)}P(k)$. 
According to Ref. \cite{cebh01}, the removal of the most connected sites leads to 
the disappearance of the links attached to them. For the network with random connections, this is equivalent to the removal of links chosen at random with the probability 
  
\begin{equation} 
\tilde{p}(p) = \sum_{k=\tilde{K}(p)+1}^\infty \!\!\!\!\!\! kP(k) \,/
\sum_{k=0}^\infty kP(k)   
\, .
\label{1}
\end{equation} 
Substituting this expression into Eq. (3) of paper \cite{cebh01} for the percolation threshold of the network having  
the degree distribution with 
the cutoff $\tilde{K}(p)$
(as one can check, this equation is valid both for the random 
removal of sites and for the random removal of links)
we obtain 

\begin{equation} 
(1 - \tilde{p}(p_c)) \sum_{k=0}^{\tilde{K}(p_c)} k^2 P(k) = 
(2 - \tilde{p}(p_c)) \sum_{k=0}^{\tilde{K}(p_c)} k P(k)
\, ,
\label{2}
\end{equation} 
so 
  
\begin{equation} 
\sum_{k=0}^{\tilde{K}(p_c)} k(k-1)P(k) = 
\sum_{k=0}^\infty kP(k)
\, .
\label{3}
\end{equation} 
Equations (\ref{2}) and (\ref{3}) are exact. 

In particular, when the degree distribution is scale-free, 
$P(k) = (1 - \delta_{k,0})k^{-\alpha}/\zeta(\alpha)$, where $\delta_{k,0}$ is the Kronecker symbol and  $\zeta(\alpha)$ is the $\zeta$-function, one can write  
$p = 1 - \sum_{k=1}^{\tilde{K}(p)} k^{-\alpha} /\zeta(\alpha)$, 
and Eq. (3) takes the form  
  
\begin{equation} 
\sum_{k=1}^{\tilde{K}(p_c)} k^{2-\alpha} = 
\zeta(\alpha-1) + \sum_{k=1}^{\tilde{K}(p_c)} k^{1-\alpha}   
\, .
\label{4}
\end{equation}   

\begin{figure}
\epsfxsize=71mm
\epsffile{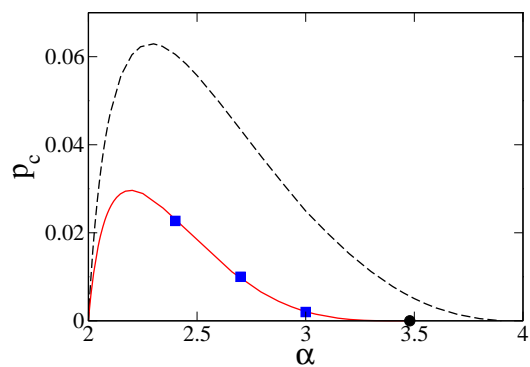}
\caption{ 
Critical fraction $p_c$  
vs. the exponent $\alpha$ for infinite scale-free networks. 
The solid line shows the exact result. The dashed line depicts the dependence 
obtained in Ref. \protect\cite{cebh01} in 
the continuum approximation. 
The circle indicates the point $\alpha^\ast=3.479\ldots$ above which $p_c=0$. 
In the continuum approximation, $\alpha^\ast=4$ \protect\cite{cebh01}.  
The squares 
represent the results of calculations and simulation in Ref. \protect\cite{cnsw00}. 
}
\label{f1}
\end{figure}

From the last two relations we obtain the explicit dependence $p_c(\alpha)$ shown by the solid line in Fig. \ref{f1}. This curve is far below the result of the continuum 
approach \cite{cebh01},    
the dashed line, which also follows from the continuum version of Eq. (3)  
(for brevity, we consider the infinite network). 

Here we have derived the exact Eq. (\ref{3}) using the heuristic arguments of Ref. \cite{cebh01} 
but it may also be obtained 
directly from the exact equations of paper \cite{cnsw00}. 
Therefore, these  
methods are equivalent. 

One should emphasize that the  
percolation threshold position 
is very sensitive to the amount of dead ends in the network. This is valid both for intentional and for random kinds of damage. 
The continuum approach   
does not properly account for the fraction of dead ends $P(1)$ and therefore has a large discrepancy with the exact results for the network with the genuine discrete degree distribution. 
\\

\noindent
S.N. Dorogovtsev$^{1,2}$ and J.F.F. Mendes$^{1}$ 

$\phantom{x}$\vspace{-11pt}

$^{1}${\small  
Universidade do Porto, 
Porto, Portugal} \\
$\phantom{x}$\hspace{5pt}$^{2}${\small A.F. Ioffe Physico-Technical Institute, 
St. Petersburg}
\\


$\phantom{x}$\vspace{11pt}

\newpage

\end{multicols}

\end{document}